\documentclass{aa}
\usepackage{epsfig}
\usepackage{rotating}
\usepackage{natbib}
\usepackage{graphicx}

\newcommand{\chan}{{\sl Chandra}}

\newcommand{\asca}{{\sl ASCA}}
\newcommand{\ascan}{{\sl Advanced Satellite for Cosmology and Astrophysics}}
\newcommand{\xmm}{{\sl XMM-Newton}}

\newcommand{\ein}{{\sl Einstein}}
\newcommand{\spitz}{{\sl Spitzer}}
\newcommand{\swift}{{\sl Swift}}
\newcommand{\uvot}{{\sl UVOT}}
\newcommand{\uvotn}{{\sl Ultraviolet Optical Telescope}}
\newcommand{\bat}{{\sl BAT}}
\newcommand{\batn}{{\sl Burst Alert Telescope}}
\newcommand{\xrtn}{{\sl X-ray Telescope}}
\newcommand{\xrt}{{\sl XRT}}
\newcommand{\mage}{{\sl Magellan}}

\newcommand{\vlt}{{\sl VLT}}
\newcommand{\hst}{{\sl HST}}
\newcommand{\hstn}{{\sl Hubble Space Telescope}}
\newcommand{\vltn}{{\sl Very Large Telescope}}
\newcommand{\naco}{{\sl NACO}}
\newcommand{\nacon}{{\sl NAos COnica}}
\newcommand{\tmass}{{\sl 2MASS}}
\newcommand{\gsc}{{\sl GSC-2}}

\newcommand{\ONEE}{{1E 1547.0$-$5408}}

\begin{document}

 \title{VLT/NACO near-infrared observations of the transient radio magnetar 1E 1547.0$-$5408\thanks{Based on observations collected at the European Southern Observatory (ESO), Paranal, Chile under  programme ID 079.D-0441(A)} }

\author{R. P. Mignani\inst{1}
 \and
 N. Rea\inst{2}
 \and 	
 V. Testa\inst{3}
 \and 	
 G.~L. Israel\inst{3}
\and
G. Marconi\inst{4}
\and
S. Mereghetti\inst{5}
\and
P. Jonker\inst{6}
\and
R. Turolla\inst{7} 
\and
R. Perna\inst{8}
\and
S. Zane\inst{1}
\and
G. Lo Curto\inst{4}
\and
S. Chaty\inst{9}
 }

\offprints{R. P. Mignani; rm2@mssl.ucl.ac.uk}
  
\institute{Mullard Space Science Laboratory, University College London, Holmbury St. Mary, Dorking, Surrey, RH5 6NT, UK
\and
University of Amsterdam, Astronomical Institute "Anton Pannekoek", Kruislaan 403, 1098~SJ, Amsterdam, The Netherlands
\and
INAF - Osservatorio Astronomico di Roma, via Frascati 33, Monte Porzio Catone, Rome, 00040, Italy 
\and
European Southern Observatory, Av. Alonso de Cordova 3107, Vitacura, Santiago, Chile
\and
INAF, Istituto di Astrofisica Spaziale, Via Bassini 15, Milan, 20133, Italy
\and
SRON Netherlands Institute for Space Research, Sorbonnelaan, 2, 3584CA, Utrecht, The Netherlands
\and
Department of Physics, University of Padua, via Marzolo 8, Padua, 35131, Italy
\and
JILA and Department of Astrophysical and Planetary Sciences, University of Colorado, 440 UCB, Boulder, 80309, USA
\and
AIM-Astrophysique Interactions Multi\'echelles (UMR 7158 CEA/CNRS/Universit\'e Paris 7 Denis Diderot), CEA Saclay, DSM/IRFU/Service d'Astrophysique, Batiment 709, L'Orme des Merisiers, FR-91 191 Gif-sur-Yvette Cedex, France.
}

\date{Received ...; accepted ...}

\abstract  {Despite about  a decade  of observations,  very  little is
known about the  optical and infrared (IR) emission  properties of the
Soft  Gamma-ray Repeaters (SGRs)  and of  the Anomalous  X-ray Pulsars
(AXPs),  the magnetar  candidates,  and about  the physical  processes
which drive their emission at these wavelengths. This is mainly due to
the limited number of identifications  achieved so far, five in total,
and  to   the  sparse  spectral  coverage   obtained  from  multi-band
optical/IR  photometry.}{Aim of  this  work is  to  search for a  likely
candidate counterpart  to the recently discovered  transient radio AXP
\ONEE.}{We performed the first  deep near-IR (NIR) observations ($K_s$
band)  of \ONEE\ with  the \vltn\  (\vlt) on  three nights  (July 8th,
12th, and August 17th), after the X-ray source rebrightening and during
the subsequent decay reported around June 2007.}{We detected four objects
within, or close to, the $3\sigma$  radio position of  \ONEE. The  faintest of
them  (object 1)  has a brightness $Ks  = 20.27  \pm 0.05$, which would yield an unabsorbed
X-ray--to--NIR flux ratio $F_{X}/F_{Ks} \sim 800 $ for \ONEE, i.e. on average lower than those derived for  other magnetars.
The
non-detection of   object 1 on the nights  of July 8th
and August 17th  only allowed us to set an upper  limit of $\Delta K_s
\sim 0.2$  on its  NIR variability, which  prevented us to  search for
correlations with  the radio  or X-ray flux.   We detected  no other
object at the radio position down to  a limit of  $K_s \sim 21.7$ (at $5 \sigma$),  computed in  our deepest \vlt\ image (July  12th). }  {From our observations we can not confidently propose a NIR counterpart to \ONEE. More NIR observations of object 1, e.g. to determine its colors and to monitor variability,  would be conclusive to  determine whether or not it can be considered a plausible candidate.}  


\titlerunning{NIR observations of the transient radio magnetar 1E\, 1547.0$-$5408}

   \maketitle

\section{Introduction} 

Since  the  discovery of  the  first  Soft  Gamma-ray Repeater  (SGR\,
0526$-$66), almost exactly  30 years ago during the  famous ``March 5th
1979'' event (Mazets  et al. 1979), five SGRs  have been discovered so
far,  the last  one (SGR\,  0501+4516)  in August  2008 (Barthelmy  et
al. 2008).   After that of SGR\,  0526$-$66 came the  discovery of the
X-ray pulsar 1E 2259+586 (Gregory  and Fahlman 1980) which, only after
a decade, was  recognized by Mereghetti and Stella  (1995) as a member
of  a  new  class  of   X-ray  sources,  the  Anomalous  X-ray  Pulsar
(AXPs). Now,  nine AXPs have  been discovered plus one  candidate (see
Mereghetti  2008 for  an  updated  review).  Both  SGRs  and AXPs  are
believed to be magnetars,  isolated neutron stars with magnetic fields
above the  quantum limit ($4.14 \times  10^{13} $ G),  as suggested by
the estimates  of their surface dipolar magnetic  field from spin-down
measurements  ($P=2-10$s; $\dot  P=10^{-11}-10^{-10}$s  s$^{-1}$).  At
variance  with  other classes  of  X-ray  pulsars,  these objects  are
thought to be  powered by energy stored in  their huge magnetic fields
(Duncan  \&  Thompson 1992;  Thompson  \&  Duncan  1995), since  their
rotational energy is  not a sufficient reservoir, and  no evidence for
accretion has been found so far.

While all magnetars  are bright in the soft X-ray band,  only a few of
them  are identified  in the  optical/near-infrared (NIR)  domain.  So
far, the only  SGR with an NIR counterpart  is SGR\, 1806$-$20 (Israel
et al.  2005; Kosugi et  al.  2005).  Among the AXPs, NIR counterparts
have been identified  for 4U\, 0142+61 (Hulleman et  al.  2000, 2004),
1E\,  1048.1$-$5937 (Wang  \& Chakrabarty  2002;Israel et  al.  2002),
XTE\, J1810$-$197  (Israel et  al.  2004; Rea  et al. 2004),  and 1E\,
2259+586 (Hulleman  et al.  2001). The  first two AXPs  have also been
detected in  the optical, where their  counterparts feature pulsations
at the X-ray period (Kern \& Martin 2002; Dhillon et al.  2008).  4U\,
0142+61 is  also the  only AXP/SGR detected  in the mid-IR  by \spitz\
(Wang et  al.  2006).  The  originally proposed NIR  identification of
1RXS\, J170849$-$400910 (Israel  et al.  2003) has now  been ruled out
by  more  recent observations  (see,  e.g.   Testa  et al.   2008  and
references  therein).  Recently, NIR  counterparts have  been proposed
for  SGR\, 1900+14 and  1E\, 1841$-$045  (Testa et  al.  2008)  on the
basis of possible  long-term variability, although the identifications
still need confirmation.

The  optical/NIR   emission  spectra   of  magnetars  are   not  fully
characterized  yet.   This is  mainly  due  to  the high  interstellar
extinction  which reduces  the spectral  coverage mostly  to  the NIR,
while lower spatial resolution mid-IR observations are hampered by the
local crowding.  Moreover, in  some cases the multi-band data-sets are
derived from  a compilation of measurements taken  at different epochs
which  sample different  source  states and,  thus,  are not  directly
comparable. Interestingly, in all  cases the counterpart fluxes fall a
few  orders of magnitude  above the  extrapolation in  the optical/NIR
domain of  the blackbody component of  the soft X-ray  spectrum. It is
not clear  whether this NIR excess is  due to a break  in the magnetar
spectrum or to the presence  of an additional source of emission which
can be  identified in  a debris  disc (Perna et  al. 2000),  like that
claimed around  the AXP  4U\, 0142+61 (Wang  et al.   2006).  However,
both for this source and  for the AXP 1E\, 1048.1$-$5937 the detection
of optical pulsations  with pulsed fractions equal to  the X-ray ones,
or larger,  and the lack of  evidence for time lags  between the X-ray
and optical pulses  (Kern \& Martin 2002; Dhillon  et al.  2008) would
argue against  the optical/NIR emission  being due to  reprocessing of
the X-ray radiation  by a disc.  A counter argument  also comes by the
observed  erratic  NIR  variability   of  XTE\,  J1810$-$197  (Camilo et al. 2007b; Testa et al. 2008), difficult to explain in the reprocessing
disc scenario.
However,  a disc  origin of  the NIR  emission is  still  considered a
possibility (see,  e.g.  Ertan et al.   2007), and the  upper limit on
the polarisation of the NIR  flux of 1E\, 1048.1$-$5937 (Israel et al., in preparation) does not provide compelling evidence for a non-thermal origin of
its NIR  emission. Interestingly, Mignani  et al.  (2007)  pointed out
that  the  high  NIR   emission  efficiency  of  magnetars  is  likely
associated with their huge magnetic fields.

One of the AXPs which still  lack deep NIR observations is \ONEE. The
source was  detected by the  \ein\ X-ray observatory (Lamb  \& Markert
1981),  re-observed  with  the  \ascan\  (\asca) by  Sugizaki  et  al.
(2001), and later on with \xmm\  and \chan. Its position at the centre
of  the candidate  supernova  remnant (SNR)  G327.24$-$0.13, its  high
X-ray--to--NIR flux ratio, its X-ray flux decay by a factor of 10 over
$\sim$ 25 years, and its soft X-ray spectrum,
prompted Gelfand \&  Gaensler (2007) to propose \ONEE\  as a candidate
AXP.  A radio pulsar (PSR  J1550$-$5418) was then discovered by Camilo
et al.  (2007a), positionally coincident with \ONEE.  The pulsar period
($P=2.069$ s) and period  derivative ($\dot P= 2.318 \times 10^{-11}$s
s$^{-1}$) yielded  a spin-down  age of 1.4  kyrs, a  rotational energy
loss $\dot E=1.0 \times 10^{35}$ erg s$^{-1}$, and a magnetic field $B
= 2.2 \times 10^{14}$  G, thus confirming the magnetar interpretation.
A  distance of  9 kpc  was derived  from the  radio  pulsar dispersion
measure (DM),  and highly ($\sim$  100\%) polarized radio  emission at
high frequency was also measured by Camilo et al.  (2008).  Monitoring
\swift\  \xrtn\  (\xrt)  observations  of  \ONEE\  performed  in  2007
June-July found  that the source underwent a  rebrightnening since its
2006 July-August minimum  and was evolving to a  new low state (Camilo
et  al.   2007a;  Halpern  et  al.  2008).   A  long,  follow-up  \xmm\
observation then led to the discovery of X-ray pulsations at the radio
period (Halpern  et al.  2008). The  lack of a  radio detection before
2006  therefore suggests  that the  radio  activity of  the source  is
intrinsically related to its  X-ray emission state.  This makes \ONEE\
the  second  transient  radio  AXP  discovered  so  far,  after  XTE\,
J1808$-$197 (Camilo  et al.  2006).   NIR observations of  \ONEE\ were
performed  by  Gelfand  \&  Gaensler  (2007) with  the  6.5  m  \mage\
telescope but they did not  pinpoint any candidate counterpart down to
$K_s \sim 17.5$.

In  this  paper  we  present   the  results  of  the  first  deep  NIR
observations  of   \ONEE\  performed  with  the   \vltn\  (\vlt).  The
observations are described in Sect. 2, together with the data analysis
and  calibration, while  the results  are presented  and  discussed in
Sect. 3 and 4, respectively.
 
 
\begin{figure*} \hbox{ \centering{ \hspace{0.8cm}
\includegraphics[height=8.cm,angle=0]{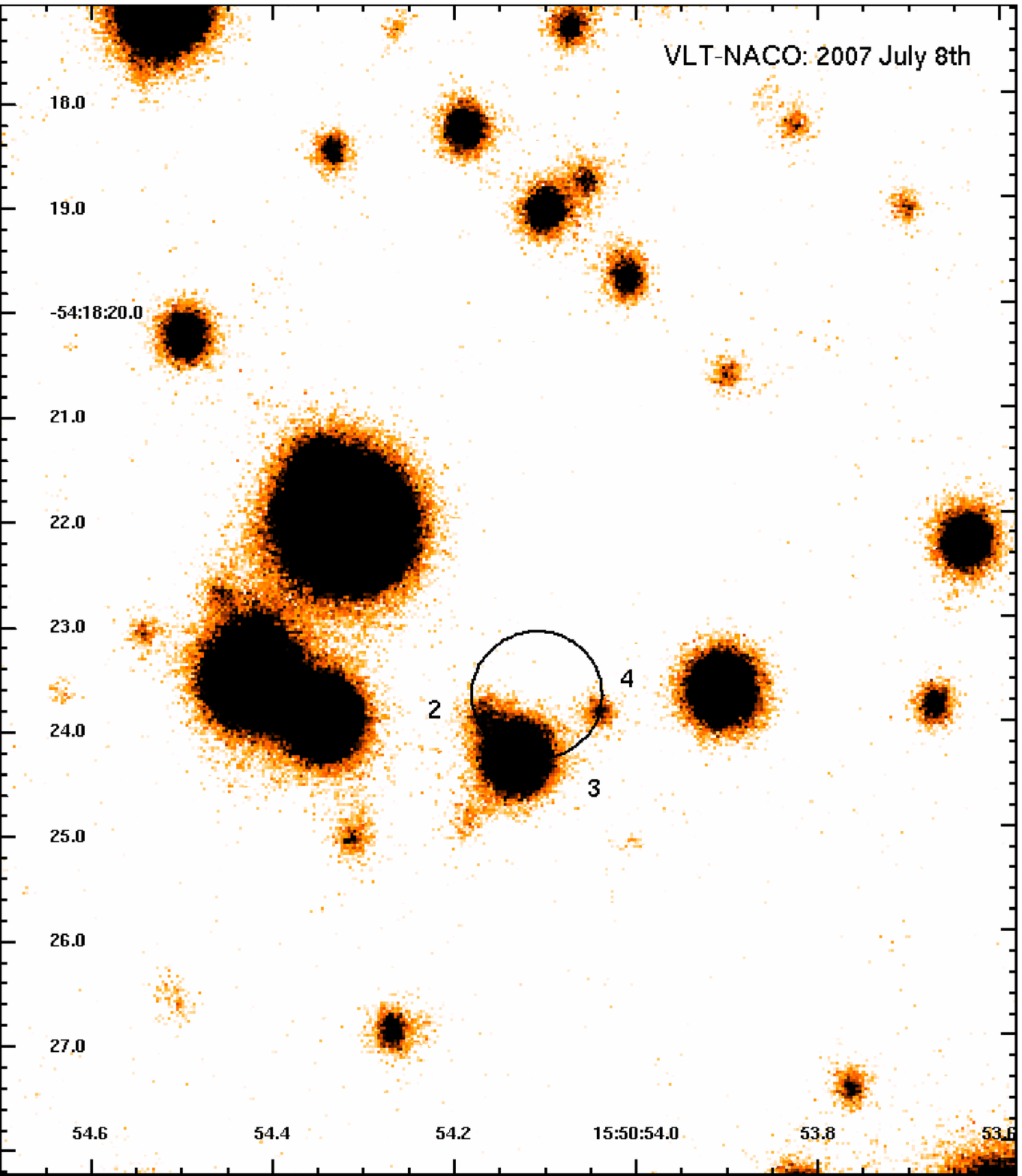}
\hspace{2cm}
\includegraphics[height=8.cm,angle=0]{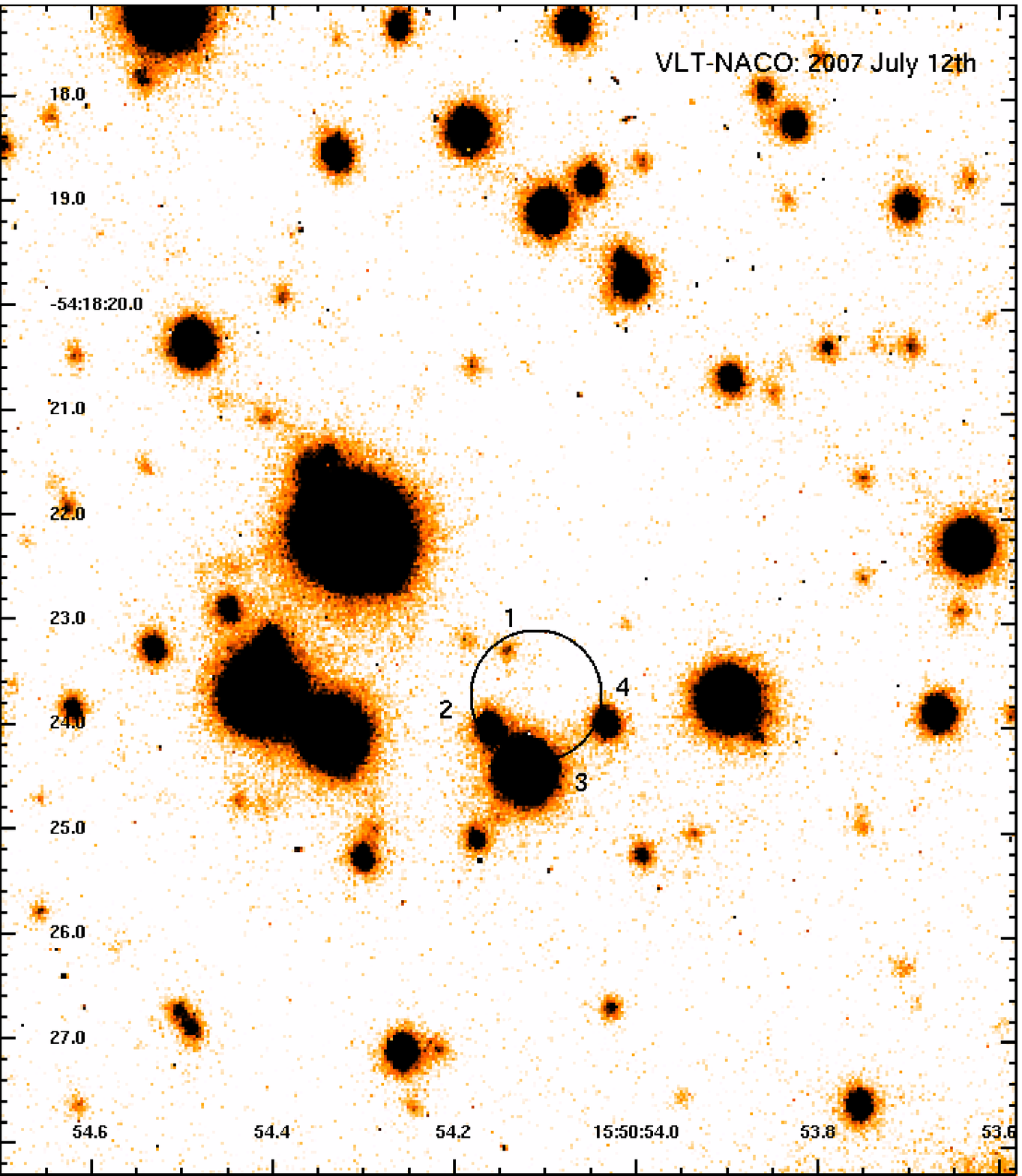}
}}
\caption{\vlt/\naco\ $K_s$-band  images of  the \ONEE\ field  taken on
2007 July 8th  (left) and 12th (right). North to the  top, East to the
left.   The radius  of  the  error circle  (0\farcs63  at 3  $\sigma$)
accounts for  the uncertainty on  the radio coordinates of  \ONEE\ and
for  the overall  error on  our astrometric  calibration  (Sect. 2.2). Objects detected within, or close to, the radio error circle are labeled.
The  bright star  south of  it (\#3;
$Ks=16.23 \pm  0.03$) is  the object detected  by Gelfand  \& Gaensler
(2007). }
\label{1e1547_fov} \end{figure*}


\section{Observations and data reduction}

\subsection{Observation description}

Thanks to  our ESO Target  of Opportunity program (ToO),  we triggered
\vlt\  observations of \ONEE\  right after  the rebrightnening  of the
X-ray source, and subsequent  decay, discovered by the \swift/\xrt\ on
2007 June 22nd (Halpern et  al.  2008).  We observed \ONEE\ in service
mode on  2007 July 8th, 12th,  and August 17th (Mignani  et al.  2008)
from the  ESO Paranal Observatory  with \nacon\ (\naco),  the adaptive
optics (AO)  NIR imager  and spectrometer mounted  at the  \vlt\ Yepun
telescope.  In  order to provide the best  combination between angular
resolution and sensitivity,  we used the S27 camera  which has a pixel
size  of  0\farcs027  and  a   field  of  view  of  $28\arcsec  \times
28\arcsec$.  As  a reference for the  AO correction we  used the \gsc\
star  S211131013236  ($V=14.3$),  located  30\arcsec\  away  from  our
target.   The  visual  ($4500-10000  ~  $ \AA)  dichroic  element  and
wavefront sensor  were used.  Unfortunately, only  observations in the
K$_s$  band were  executed as  part of  our program  due  to telescope
scheduling constraints.  To allow  for subtraction of the variable NIR
sky  background,  we split  each  integration  in  sequences of  short
dithered exposures with detector integration times (DIT) of 40 s and 4
exposures  (NDIT) along each  point of  the random  dithering pattern.
The observations log is summarised in Table \ref{data}.  In each night
the target  was observed close  to zenith, with airmass  always better
than 1.2.   Unfortunately, only for  the July 12th  observation seeing
conditions were  good enough  (i.e.  below 0\farcs8)  to allow  for an
optimized use of  the AO, while for the other  nights the seeing never
went below  1\farcs0.  In particular,  the observation of  August 17th
was affected by a seeing of $\sim 1\farcs6$. Because of its much lower
image quality and  shallower flux limit with respect  to the others we
did not use this observation in the following analysis.  Moreover, the
observation  of July  8th was  affected by  a strong  wind,  above the
pointing limit,  and {\bf possibly} by the  presence of clouds.  Sky  conditions were
photometric  for  the July  12th  night  only.   Daytime (darks,  lamp
flat--fields)   and   night    time   calibration   frames   (twilight
flat--fields),  were taken  daily as  part of  the  \naco\ calibration
plan.   Standard stars  from the  Persson et  al.  (1998)  fields were
observed for photometric calibration at the beginning of each night.

\begin{table}[h]
\begin{center}
\caption{Log of  the \vlt/\naco\ Ks-band observations of the \ONEE\ field. 
  Columns report the observing epoch and Modified Julian Date (MJD), the total exposure time T = DIT $\times$ NDIT $\times$ N, where N is the number of nodes of the dithering pattern, and the 
average seeing and airmass during the observations. }
\begin{tabular}{ccccc} \\ \hline 
 Epoch    & MJD &  Exp. Time  & Seeing  & Airmass	\\
 (yyyy-mm-dd) &        & (s)    &  ($\arcsec$)   & 	\\ \hline
 2007-07-08   &  54289 & 1440   &  1.18  &   1.16  \\
 2007-07-12   &  54293 & 1440   &  0.55  &   1.17  \\
 2007-08-17   &  54329 & 1800   &  1.60  &   1.19  \\  \hline
\end{tabular}
\label{data}
\end{center}
\end{table}

\subsection{Data reduction and calibration}

The \vlt\  data were processed  through the ESO \naco\  data reduction
pipeline\footnote{www.eso.org/observing/dfo/quality/NACO/pipeline}.
Science  frames  were  reduced  with  the  produced  master  dark  and
flat--field frames and combined  to correct for the exposure dithering
and to produce cosmic-ray free and sky-subtracted images.  We computed
the astrometric  calibration using the coordinates and  positions of 6
stars selected  from the \tmass\  catalogue (Skrutskie et  al.  2006).
The  pixel coordinates  of the  \tmass\ stars  (all non  saturated and
evenly  distributed  in the  field)  were  measured  by fitting  their
intensity profiles  with a Gaussian function using  the {\tt Graphical
Astronomy       and      Image       Analysis}       ({\tt      GAIA})
tool\footnote{star-www.dur.ac.uk/~pdraper/gaia/gaia.html}.  The fit to
celestial coordinates  was computed  using the {\sl  Starlink} package
{\tt  ASTROM}\footnote{star-www.rl.ac.uk/Software/software.htm}.   The
rms  of the  astrometric fit  residuals was  $\sigma_{\rm fit}\approx$
0\farcs05  in   the  radial  direction.   We   estimated  the  overall
uncertainty of  our astrometric solution  by adding in  quadrature the
$\sigma_{\rm   fit}$   and   the   uncertainty  $\sigma_{\rm   tr}   =
\sqrt{3/N_s}\,\sigma_{\rm 2MASS}=0\farcs14$ with which we can register
our field on the \tmass\  reference frame.  In this case, $\sigma_{\rm
2MASS}=0\farcs2$  is a conservative  mean random  radial error  of the
\tmass\ coordinates (Skrutskie et al.  2006), $N_s=6$ is the number of
reference stars,  and $\sqrt{3}$ accounts  for the free  parameters in
the astrometric fit (e.g., Lattanzi et al.\ 1997).  The uncertainty on
the  reference star centroids  is below  0\farcs01 and  was neglected.
This  yields an  overall  uncertainty of  our  absolute astrometry  of
0\farcs15.  Since the \tmass\  astrometry is tied to the International
Celestial Reference Frame (ICRF) with  a $\sim$ 15 mas accuracy, there
is no significant shift between the radio and the NIR coordinates.

As  a  reference  for  the  photometric  calibration  we  assumed  the
airmass-corrected  zero  points   provided  by  the  \naco\  pipeline,
computed through  fixed aperture photometry.  However,  the zero point
of July 8th shows a  deviation of $\sim0.5$ magnitudes with respect to
the             mean             of            the             trended
values\footnote{www.eso.org/observing/dfo/quality/NACO/qc/zp\_qc1.html}
which has to  be ascribed to the non-photometric  conditions.  We thus
used   the  zero   point  measured   on   the  night   of  July   12th
(23.24$\pm$0.02).    We   then   performed  a   relative   photometric
calibration of the July 8th data by computing a linear fit between the
measured  instrumental magnitudes  of  $\sim$ 50 sufficiently  bright
field stars detected  in the field.  For a  direct comparison with the
zero point computed by the  \naco\ pipeline, we computed object fluxes
through  aperture  photometry  and   we  accounted  for  the  aperture
correction.   We used  the {\sl  IRAF}  version of  the {\sl  Daophot}
package  (Stetson  1992).  The  fit  yielded  an  rms of  $\sim  0.12$
magnitudes which  we assume  as the systematic  error of  our relative
photometry calibration.

\section{Results}

As a reference  for the \ONEE\ position we  used the radio coordinates
from  Camilo  et  al.   (2007a):  $\alpha_{J2000}=15^h  50^m  54.11^s$,
$\delta_{J2000}=  -54^\circ   18\arcmin  23\farcs7$,  which   have  an
absolute  accuracy of  0.01$^s$ and  0\farcs1 in  right  ascension and
declination, respectively.  The computed $3 \sigma$ radio error circle
of \ONEE\ is shown in Figure  1, registered on the July 8th (left) and
12th  (right) \naco\  $K_s$-band  images.  Four  objects are  detected
within, or close to, the {\bf $3\sigma$} radio  error circle and are labeled  in Figure 1.
The brightest one (\#3) is the  object detected in the \mage\ image of
Gelfand \&  Gaensler (2007).  Three  more objects are now  detected in
our image, one of which (\#2) is very close to the Gelfand \& Gaensler
star and  better resolved in  the July 12th  image thanks to  its much
better  image  quality. The  faintest  of  the  four objects (\#1),   is detected in
the July 12th image  only.  This is by far the deepest  of the two, as
shown  by the  difference in  the number  of objects  detected  in the
field.  The angular distance of objects \#1-4 from the nominal radio position is reported in the second column of Table 1. We note that only objects \#1 and 2 are detected within the $3 \sigma$ radio error circle. Although the not so tight  positional coincidence argues against an association with the X-ray source, it does not formally rule it out. 

We measured the magnitudes  of objects  \#1-4 through  PSF photometry
with {\sl  Daophot}, to better  resolve the emission from  object \#2
which is  partially blended with object  \#3 even  in the  best image-quality
data of July  12th.  The results of our photometry  for the two nights
are shown in Table 2, for  comparison.  We note that the measured flux
of object  \#3 ($K_s =  16.23 \pm 0.03$)  is slightly lower  than that
reported  by Gelfand  \&  Gaensler (2007)  who  gave $K_s  = 15.9  \pm
0.2$. This is most likely due to the fact that objects \#2 and \#3 are
not resolved  in their lower  resolution images (see their  Figure 3),
which yields a  flux overestimation for object \#3.   For objects \#2,
as well as for 3  and   4,  the  measured  magnitudes  are   consistent  within  the
statistical  errors  and  the  accuracy  of  our  relative  photometry
recalibration (see Sect.  2.2 and  caption to Table 2).  Thus, none of
them is variable on the time span of our observations.  For object \#1
($K_s = 20.27 \pm 0.05$)  no direct variablity check is possible since
it  was not  detected in  the  July 8th  image.  Since  the night  was
non-photometric, its  limiting magnitude estimate  is quite uncertain.
In particular,  the residuals of our  relative photometry calibrations
show  a relatively  large scatter  of $\sim  0.2$ magnitudes  for $K_s
\approx 20$,  which we assume as  the random error  affecting the July
8th  photometry at  this flux  level.  Thus,  accounting for  the 0.12
magnitudes rms of  the recalibration fit, our estimate  is affected by
an  overall uncertainty  of $\sim  0.25$ magnitudes,  which translates
into a  $5 \sigma$ limiting magnitude  of $Ks \sim 20.1$  for the July
8th image.  This  implies an upper limit on the object \#1 variability of
$\Delta  K_s \sim  0.2$, accounting  for the  error of  the  July 12th
measurement.  Since none  of the objects close to  the \ONEE\  position
is variable, with  the possible exception of \#1,  in the following we
use their magnitudes measured on the July 12th image.  We measured its
$5 \sigma$ limiting magnitude to be $Ks \sim 21.7$.

\begin{table}
\begin{center}
\caption{$K_s$-band magnitudes of  the candidate counterparts shown in
Figure 1,  measured for the  two nights.  The second column gives the source angular distance  from the nominal radio position of \ONEE. Only statistical  errors ($1
\sigma$) are listed.  In all cases, the actual photometry error has to
account for the zero point  error (0.02 magnitudes) and, for the night
of July 8th, also for the rms of our relative photometry recalibration
(0.12 magnitudes). }
\begin{tabular}{cccc} \\ \hline
 Source ID            & $\Delta ~r$ & 2007-07-08 &  2007-07-12  \\ \hline
 1 &  0\farcs50 & $\ge$ 20.1       & 20.27 $\pm$ 0.05  \\
 2 &  0\farcs54 & 18.65 $\pm$ 0.08 & 18.51 $\pm$ 0.03  \\
 3 &  0\farcs76 & 16.22 $\pm$ 0.05 & 16.23 $\pm$ 0.03  \\
 4 &  0\farcs74 & 18.56 $\pm$ 0.06 & 18.54 $\pm$ 0.03  \\ \hline
\end{tabular}
\label{phot}
\end{center}
\end{table}

\section{Discussion}

The lack of color information  does not enable us to  determine  which
of the objects detected within the $3\sigma$ radio error circle (Figure 1)
might be a   plausible candidate counterpart to  \ONEE.  While object  \#2, as well as 3 and 4,  is
clearly non-variable between 2007 July  8th and 12th, object \#1 might
still feature a tiny variability ($\Delta K_s \le 0.2$). From  the observed
X-ray  flux decay rate  of \ONEE\  (Halpern et  al.  2008), this would not rule out the variation expected if the NIR flux were to scale at the same rate of the X-ray  one.  
 Thus, we can  not use the comparison  between NIR and
X-ray variability  to investigate  a possible identification,  as done
for other magnetars.

We  then considered  the  unabsorbed X--ray-to-NIR  flux  ratio as  an
alternative criterion.   The X-ray  spectrum  of \ONEE\
measured by \swift\  on 2007 July 9th, 10th, and  16th (Halpern et al.
2008), i.e.   few days before and  after our deeper NIR  observation, can be
fit   by   an   absorbed    blackbody   plus   power-law   ($N_{H}   =
3.2\times10^{22}$\,cm$^{-2}$,  kT=  0.5\,keV  and $\Gamma=2.93$),  two
blackbodies  ($N_{H} = 3.0\times10^{22}$\,cm$^{-2}$,  kT$_1$= 0.5\,keV
and  kT$_2$=  1.3\,keV),  or  a resonant  cyclotron  scattering  model
($N_{H}  = 3.0\times10^{22}$\,cm$^{-2}$,  kT=  0.5\,keV, $\tau=1$  and
$\beta=0.24$; Rea et  al.  2008a)\footnote{The $N_{H}$ values reported
here are slightly different from  those in Rea et al.~(2008a) because,
to be  consistent with  Halpern et al.~(2008),  we computed  the N$_H$
assuming  the solar  abundance values  of Anders  \&  Grevesse (1989),
while in Rea et al. (2008a)  the more updated ones from Lodders (2003)
has  been used.  Other  spectral parameters  are not  affected.}.   As a reference, we plotted in Figure 2 the best fit to  the \swift/\xrt\  X-ray spectrum computed using  two  blackbodies  and  a resonant  cyclotron scattering (RCS) model, together with the dereddened
$K_s$-band flux of objects  \#1-4  and the  $K_s$-band flux upper limit. 
We computed the  interstellar extinction correction  in the  $K_s$ band  using the
hydrogen column density $N_{H}$ derived  from the spectral fits to the
\xrt\ data and applying the relation of Predhel \& Schmitt (1995) with
the extinction  coefficients of Fitzpatrick (1999).  

Both the  two blackbodies and the RCS spectral models yield an unabsorbed X-ray flux $F_{X} \sim 5.8 \times 10^{-12}$ erg cm$^{-2}$  s$^{-1}$ in the  2-10 keV energy range.    The derived NIR flux  $F_{Ks}$ of object \#1 would imply an unabsorbed X-ray--to--NIR flux ratio $F_{X}/F_{Ks} \sim 800$  for \ONEE, where $F_{Ks}$ is integrated over the $K_s$ band.  
 We consistently recomputed the unabsorbed $F_{X}/F_{Ks} $ ratios for all magnetars with a secured NIR counterpart and  we found a range of values which are  on average a factor of 10 higher, or more.  
However, we note that there are several caveats in comparing  $F_{X}/F_{Ks}$ ratios  which, for a given source, are not always derived from contemporary X--ray and NIR observations and, for different sources, sample different X-ray states.  Moreover,  the computed X-ray fluxes
can vary significantly, being much higher or lower  depending on the assumed  spectral model, while different NIR spectra can yield relatively larger $F_{Ks}$ and, consequently, lower $F_{X}/F_{Ks}$.  Thus, it is not obvious to firmly determine whether or not object  \#1 can be considered a plausible candidate counterpart to \ONEE. 
 
Given the crowding  of the field, we also have  to estimate the probability
of  a  chance  coincidence  between  the  position  of   object  \#1 and that of \ONEE.  The chance coincidence probability can 
be estimated as $P=1-\exp(-\pi\rho r^2)$, where $r$ is the radius of
the  $3 \sigma$  radio  error  circle (0\farcs63)  and  $\rho$ is  the
density of stellar  objects in the \naco\ field  of view.  We computed
$\rho$ from  the object catalogue  obtained by running  {\sl Daophot},
selecting  star-like  objects  (ellipticity $e<0.2$)  with  magnitudes
comparable to that of object \#1
and  detected  with at  least  $5\sigma$  significance  to filter  out
background fluctuations.  In order  to avoid systematic effects due to
the  degradation  of the  \naco\  PSF  at  large off-axis  angles,  we
computed $P$ on the  inner $13\farcs5 \times 13\farcs5$ region centred
on the source position.   We found $\rho \sim0.03$ arcsec$^{-2}$ which
gives a chance coincidence probability $P\sim 0.04$, certainly not low
enough to  firmly claim,  on the basis  of the  positional coincidence
only, an association of object  \#1 with \ONEE. 

 
\begin{figure} 
\centering{
\includegraphics[height=8.5cm,angle=270]{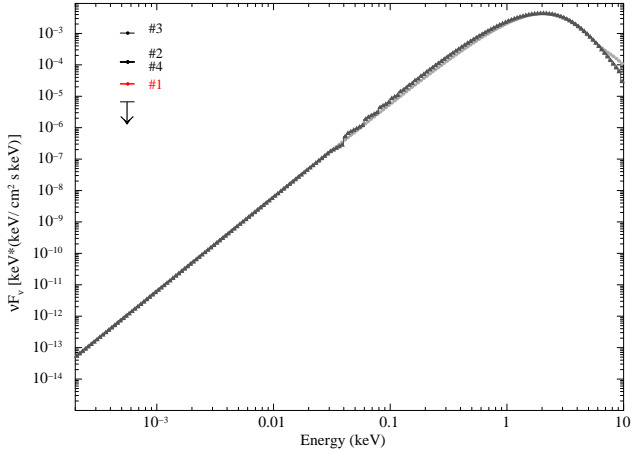}}
\caption{Fits to  the \swift/\xrt\  X-ray spectrum of  \ONEE.
To increase the
signal--to--noise ratio, we  merged the 2007 July 9th,  10th, and 16th
datasets. Light  and  dark  grey
curves represent  the  fit with  two  blackbodies  and  a resonant  cyclotron
scattering (RCS) model, respectively (see  text for details). We did not plot the blackbody plus power-law model since the latter  component  diverges  when  extrapolated  to  the
optical/NIR  domain. The dereddened
$K_s$-band flux of object \#1 (red)  and 2-4  (black)  and the  $K_s$-band flux upper limit
derived from   the  July   12th observation are plotted. }
\label{1e1547_vFv} 
\end{figure}

\section{Conclusions}

Using \naco\ at the \vlt\ we performed the deepest NIR observations so
far of the radio transient AXP  \ONEE.  We discovered a faint object (\#1; $Ks =
20.27 \pm  0.05$) within  $3 \sigma$ from  the source  radio position, 
which would imply an $F_{X}/F_{Ks}$ ratio of $ \sim 800$, i.e. on average lower than those derived for other magnetars. Thus,  we can not confidently propose object \#1  as the NIR counterpart to \ONEE.   More  observations, to  study its
colors  and to search for NIR variability
correlated  with the  radio or  X-ray  one,  would be conclusive to settle this issue.
 We note that at  the  time  of  writing  of  this
manuscript a series of bursts from \ONEE\ were detected by the \swift\
\batn\  (\bat) on  October 3  2008  (Krimm et  al.  2008;  Rea et  al.
2008b; Israel et al. in preparation), when  the source was at  the edge of the  visibility window of
ground-based  telescopes  and  right  before  the  \hstn\  (\hst)  was
switched to safe mode.
Follow-up \swift\ \uvotn\ (\uvot) observations (Holland \& Krimm 2008)
did not unveil any new candidate counterpart.  On the other hand, if
object \#1  is an  unrelated field object,  our
observations would set a limit of $K_s \sim 21.7$ on the brightness of
the undetected \ONEE\ counterpart,  about 3 magnitudes deeper than the
previously reported limit.

\begin{acknowledgements}
 RPM  acknowledges STFC  for support  through a  Rolling Grant.  NR is
 supported  by  an  NWO  Veni  Fellowship,  and  thanks  B.  Gaensler,
 J.D. Gelfand ,and F. Camilo  for useful discussions on this source. SZ
 acknowledges STFC for support through an Advanced Fellowship.

\end{acknowledgements}

\end{document}